\documentclass[preprint2,twoside]{hwo}

\usepackage{array}

\newcommand{\hst}{{\em HST}\/}

\newcommand{\GALEX}{{\em GALEX}\/}

\newcommand{\lya}{Ly$\alpha$}

\newcommand{\lyc}{LyC}

\newcommand{\fec}{$f^e_{LyC}$}
\newcommand{\fen}{$f^e_{900}$}
\newcommand{\flunit}{erg\,s$^{-1}$\,cm$^{-2}$\,\AA$^{-1}$}
\newcommand{\lunit}{erg\,s$^{-1}$\,\AA$^{-1}$}
\newcommand{\PreserveBackslash}[1]{\let\temp=\\#1\let=\temp}
\let\pb

\bibliographystyle{aasjournal-rev}

\input{hwo.h}

\begin{document}

\title{\textbf{\LARGE Evolution of the Ionizing Photon Luminosity Function}}
\author {\textbf{\large Stephan R. McCandliss,$^{1}$ Swara Ravindranath,$^{2,3}$ Sangeeta Malhotra,$^2$ Chris Packham,$^4$ Sophia Flury,$^5$ Alexandra Le Reste,$^6$ Allison Strom,$^{7}$ Marc Postman,$^8$ John O’Meara$^9$
 }}
\affil{$^1$\small\it Johns Hopkins University, Department of Physics \& Astronomy, Center for Astrophysical Sciences, 3400 North Charles Street, Baltimore, MD, USA, 21218}

\affil{$^2$\small\it Astrophysics Science Division, NASA Goddard Space Flight Center, 8800 Greenbelt Road, Greenbelt, MD 20771, USA}

\affil{$^3$\small\it Center for Research and Exploration in Space Science and Technology II, Department of Physics, Catholic University of America, 620 Michigan Ave N.E., Washington, DC 20064, USA}

\affil{$^4$\small\it Department of Physics and Astronomy, The University of Texas at San Antonio, 1 UTSA Circle, San Antonio, TX 78249-0600, USA}

\affil{$^5$\small\it Institute for Astronomy, University of Edinburgh, Royal Observatory, Edinburgh, EH9 3HJ, UK}

\affil{$^6$\small\it Minnesota Institute for Astrophysics, University of Minnesota, 116 Church Street SE, Minneapolis, MN 55455, USA}

\affil{$^7$\small\it Department of Physics and Astronomy, Northwestern University, 2145 Sheridan Road, Evanston, IL 60208, USA}

\affil{$^8$\small\it Space Telescope Science Institute, 3700 San Martin Drive, Baltimore, MD 21218, USA}

\affil{$^9$\small\it W.M. Keck Observatory, 65-1120 Mamalahoa Highway, Kamuela, HI 96743, USA}



\author{\footnotesize{\bf Endorsed by:}
Danielle Berg (University of Texas Austin), Cody Carr (Zhejiang University), Hsiao-Wen Chen (University of Chicago), Annalisa Citro (University of Minnesota), Michael Davis (Southwest Research Institute), Lukas Furtak (Ben-Gurion University of the Negev), Alaina Henry, (STScI), Anton Koekemoer (STScI), Eunjeong Lee (EisKosmos (CROASAEN), Inc), Laura Pentericci (INAF-Astronomical Observatory of Rome), Marc Rafelski (STScI), Julia Roman-Duval (STScI), Sylvain Veilleux (University of Maryland, College Park)
}

\begin{abstract}
Counting the number and brightness of ionizing radiation sources out to a redshift of $z \sim$ 1.2 will revolutionize our understanding of how the ionizing background is created and sustained by the embedded growth of meta-galactic structures.  The sheer number of sparsely separated targets required to efficiently construct redshift binned luminosity functions is industrial in scale, driving the need for low spectral resolution multi-object spectroscopy (MOS) with a short wavelength cut-off $\sim$ 1000 \AA, a sensitivity in the far-UV to better than 30 abmag, and an instantaneous field-of-view $\sim$ (2\arcmin)$^2$.  A MOS on Habitable Worlds Observatory is the only instrument that could conceivably carry out such an ambitious observing program. This program will quantify how much of the ionizing radiation  produced by galaxies is attenuated by intervening neutral H, He and dust, and how much escapes to maintain the universe in a mostly ionized state.
\\ \\
\end{abstract}

\vspace{2cm}

\section{Science Goals}

How did the Universe become reionized, how is the ionization sustained over cosmic time, and  
what are the sources and sinks of the meta-galactic ionizing background? These questions are part and parcel of the topics prioritize in the Astro2020 Decadal Survey: 

\begin{itemize}
\item[D-Q1~] – How did the intergalactic medium and the first sources of radiation evolve from cosmic dawn through the epoch of reionization?
\item[D-Q1a] – Detailed Thermal History of the Intergalactic Medium and the Topology of Reionization
\item[D-Q1b] – Production of Ionizing Photons and Their Escape into the Intergalactic Medium
\item[D-Q1c] – Properties of the First Stars, Galaxies, and Black Holes
\end{itemize}

\subsection{Overview}
We know the universe is mostly ionized. How this happens is a mystery.  

Following the initial creation of the first atoms, of which hydrogen is the most abundant \citep{Gamow1946, Alpher1948, Gamow1948}, the universe enters a phase of smooth expansion and gradual cooling.  The cooling causes the hydrogen to coalesce into dense clouds of gas that collapse under the weight of gravity to form stars, some of which are very massive \citep{Silk1983}. Such massive stars emit radiation that is so energetic that any nearby hydrogen in the surrounding clouds is ripped into its component electrons and protons, a process known as ionization. This process proceeds slowly at first, in tiny isolated ionized pockets filled with energetic radiation bounded by surrounding (neutral) hydrogen \citep{Kashlinsky1983}.  Eventually, neighboring  pockets of radiation filled with ionized gas merge to create structures of dense hydrogen surrounded by free streaming ionizing radiation \citep{Gnedin1997}.  

\begin{figure}[t]
\includegraphics[page=1,width=0.5\textwidth, viewport = 1.in .8in 11in 7.5in, clip= ]{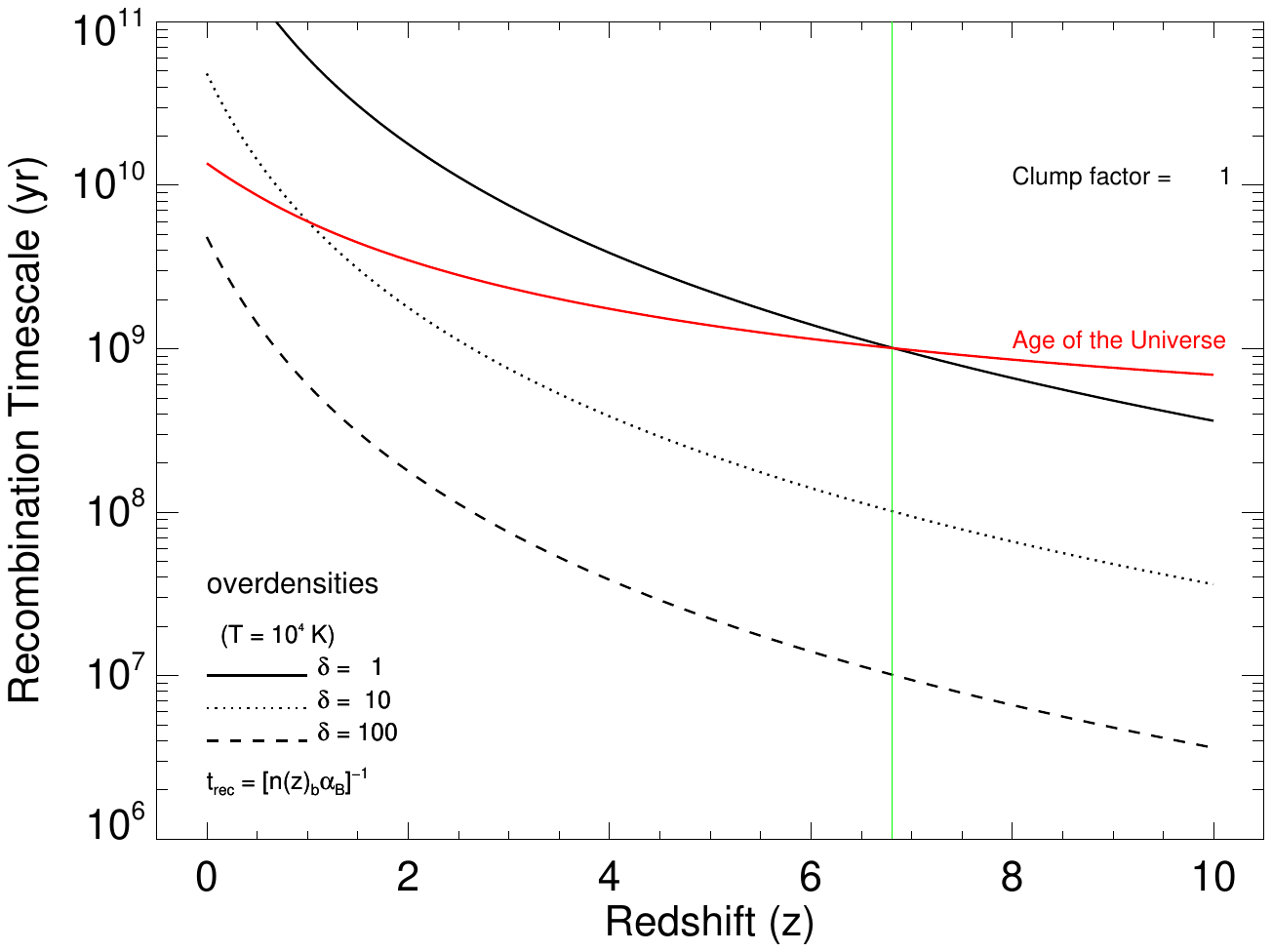}
\includegraphics[page=3,width=0.5\textwidth, viewport = 1.in .8in 11in 7.5in, clip= ]{RecombineRate_Clumps.pdf}
\caption{Top -- Recombination timescales compared to the age of the universe as a function of redshift for hydrogen over-densities of $\delta$ = 1, 10, and 100, under the assumption of Case B recombination at a temperature of 10$^4$ K.  Bottom -- Baryon densities for the same over-dense regions.}
\label{reion}
\end{figure}

The mystery is that the dense neutral hydrogen quickly becomes ionized, yet a dense ionized gas of electrons and protons (known as a plasma) quickly recombines back to neutral hydrogen on a timescale of, 
\begin{equation}
t_r  \approx (n_h \alpha_B)^{-1},
\end{equation}
where $n_h$ is the hydrogen density, and $\alpha_B$ = 2.6 $\times$ 10$^{-13}$ cm$^3$ s$^{-1}$ assuming case-B recombination at T=10$^4$K.  In the top panel of Figure~\ref{reion} we compare 
recombination timescales as function of redshift for various over-densities with respect to the mean baryon density given by, 
\begin{equation}
n_{h} = \frac{3 H_{0}^{2}\Omega_{b}}{8 \pi G M_{H}} (1+z)^3, \label{deneq}
\end{equation}
where  $H_{0}$ = 70 km s$^{-1}$ Mpc$^{-1}$ and $\Omega_{b}$ = 0.046.  
Given this simple assumption, we see that the recombination timescale naturally exceeds the age of the universe at a redshift $\sim$ 7 for unity over-densities.  Once ionized, such regions will maintain their ionization state forever.  

However, over-dense regions have recombination timescales that are  shorter, requiring a persistent source of ionizing radiation to maintain the plasma state. This is because dense neutral hydrogen is extremely efficient in absorbing ionizing radiation and when it does it produces a recombination cascade; a fluorescence process that converts the ionizing radiation into lower energy free-bound and bound-bound transitions of hydrogen. In the process the radiation field progressively becomes less energetic and plasma state is quenched unless there is a sustaining source of ionizing radiation to maintain the ionized state in these over-dense regions and the (mostly) ionized universe.  For reference, the bottom panel of Figure~\ref{reion} shows how the baryon density of the expanding universe varies with redshift (Eq~\ref{deneq}). 

Over time this cascade of structure formation through coalescence, collapse, ionization and recombination builds the galaxies and voids that we observe today.  They were seeded in the ancient universe we see streaming to us from cosmological distances.  By measuring the amount of energetic ionizing radiation in these structures over cosmic time we gain insight into how these structures form, persist and whether they harbor niches as unique as our own galaxy and our place within it.

We seek to measure the amount of escaping ionizing radiation from a large number of galaxies of diverse sizes and environments to create an ionizing radiation ``luminosity function'' \citep{Schechter1976} that characterizes the number density of galaxies in a finite volume (redshift interval) per unit luminosity bin.  The statistical heft offered by such a sample enables tests against theories of structure formation that predict the degree of sustained ionization in the present-day universe.  It also enables an investigation into the search for redshift (low-z) analogs to the galaxies that ionized the universe during the epoch of reionization around a redshift of $z \sim$ 7.

\begin{figure*}
    \centering
   \includegraphics[width=0.4\linewidth]{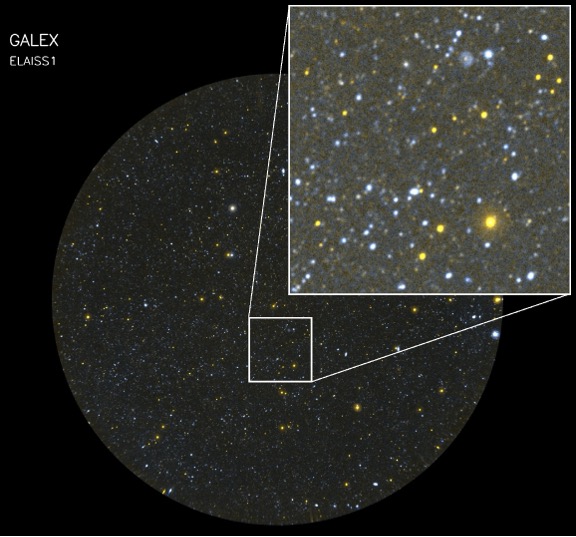}
   \includegraphics[width=0.59\linewidth,viewport=0in .22in 6.75in 4.6in,clip=]{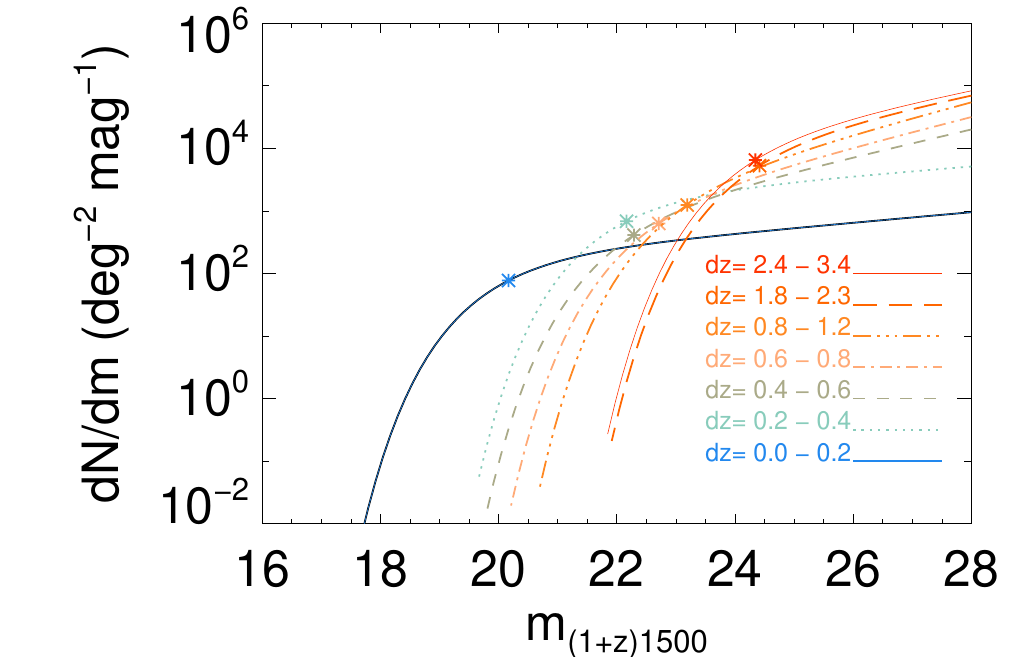}
    \caption{ Left -- \GALEX\ image of ELAIS (European Large Area ISO- Survey) field showing the far-UV sky is filled with sparsely separated  star-forming galaxies.  Right -- areal density at rest frame 1500 \AA\ for a variety of redshift intervals derived from \citet{Arnouts2005}. The asterisk marks the ab-magnitude of the “median galaxy” brightness often referred to as an ``$L^*$ Galaxy''}
    \label{galex}
    \vspace{-.2in}
\end{figure*}

\section{Science Objectives}

\begin{itemize}
\item {\bf Directly detect ionizing radiation leaking from a statistically representative sample of galaxies at low $z$  where the IGM is relatively transparent (out to $z \sim$ 1.2). }
\item {\bf Measure the spectral energy distribution (SED) of the escaping ionizing radiation and its evolution.}
\end{itemize}

Counting the number of galaxies within a magnitude range is a time honored way to understand the evolution of galaxies as a function of redshift.  \GALEX\ showed that there exists a large number of far-UV emitting sources (Figure~\ref{galex})  and characterized the Luminosity Function (LF) of these galaxies over the redshift range 0.2 $< z <$ 3.4 at a rest frame wavelength of 1500 \AA\ (Figure~\ref{galex} - Right). 

\begin{figure*}[t]
\centering
\includegraphics[width=.88\textwidth, viewport=1in .5in 10in 5.5in]{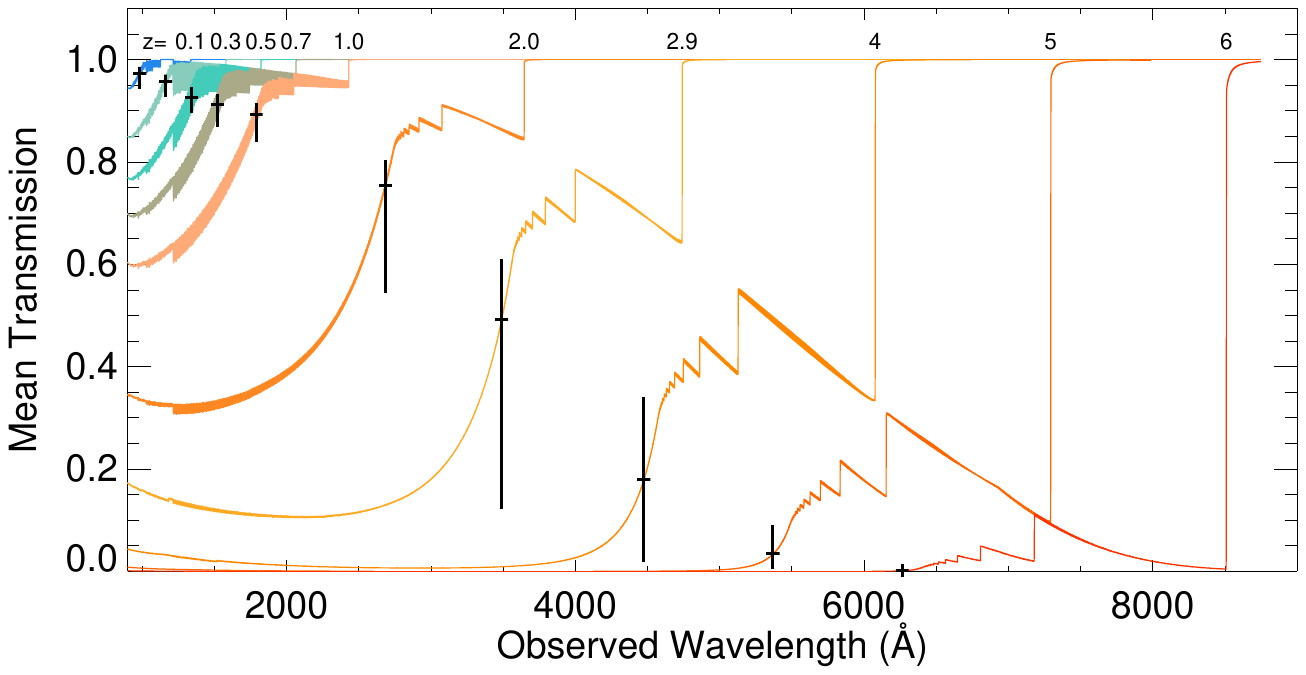}
\vspace{-.15in}
\caption{\small Mean transmission of the universe as a function of wavelength and redshift from 0.1 $\le z \le$ 6 from following \citet{McCandliss2017} and \citet{Inoue2014}.  The discrete edges are produced by the Lyman series lines smeared out over the redshift interval from z to z=0. The crosses mark the mean transmission at the photoionizaton edge in the observer frame at the indicated redshift.  The vertical height of the cross indicates the level of variation from the mean that can be expected along any given random line of sight as estimated by \citet{Inoue2008}. The large sharp drop for each curve marks the progressive increase of opacity at \lya, going completely black by a redshift of 6. The mean transmission makes detection of \lyc\ above a $z \gtrsim$ 2 problematic. } \label{meantrans}
\vspace{-.1in}
\end{figure*}

We propose a similar but much deeper program with HWO, concentrating on the very faint rest frame bandpass below the ionization edge of hydrogen (911.8 \AA), known as the Lyman Continuum (LyC).  Our goal is to carryout a statistically representative spectroscopic investigation of \lyc\ escape from all types of far-UV emitting galaxies out to the practical redshift limit of $z \la$ 1.2, as imposed by the mean opacity of the universe (Figure~\ref{meantrans}).

The recently selected UVEX with its (3\fd5)$^2$ Field-of-View (FOV) holds the promise of extending the faint end of the 1500$(z+1)$ \AA\ LF $\sim$ a magnitude deeper, thereby expanding the potential target pool and providing redshift distance indicators. It may be able to provide far-UV photometry in the rest frame LyC beyond $z \sim$ 0.7, however, it will lack the sensitivity to measure the luminosity below 900$(z+1)$ \AA\ in the statistically significant number of objects envisioned for this science case. 

Achievement of such a goal was once doubted \citep{Fernandez-Soto2003}, but has been bolstered by the remarkable success over the past decade in the direct detection of ionizing radiation escaping from compact star-forming galaxies.
 
The Low z Lyman Continuum Survey plus (LzLCS+), a combination of archival data from \citep{Izotov2016a, Izotov2016b, Izotov2018a, Izotov2018b, Wang2019, Flury2022a, Flury2022b},\footnote{The bulk of which was obtained with the efficient ``gapless'' CENWAV800 setting for the G140L mode of the Cosmic Origins Spectrograph on \hst\ \citep{Redwine2016}.} has found satistically significant detections of escape  at $\approx$ 900 \AA\ in more than half the sample of 89 objects in the redshift range 0.24 $< z <$ 0.43; some as high as \fen $<$ 60.  It is now possible to contemplate the acquisition of a robust data set to detect the dominant producers of ionization radiation, and understand how the metagalactic ionizing background (MIB) is sustained over cosmic time. 

The faintness of the ionizing flux drives us to propose low spectral resolution modes, however, ancillary observations with high spectral resolution longward of 912 \AA, where the far-UV fluxes are significantly brighter, will also allow exploration of how  metallicity, geometry, and the kinematics of galactic outflow are related to the fraction of ionizing radiation that can escape.

The emphasis on low redshift observations as the laboratory for investigating the physical processes of LyC escape is driven by the mean opacity of the universe to ionizing radiation as depicted in Figure~\ref{meantrans}.  The opacity is driven by the overlap of neutral hydrogen clouds scattered through the universe, known as the Lyman forest, which becomes progressively thicker toward higher redshift.   The figure shows the result of integrating Equation~\ref{eqtrans}, the distribution function of intergalactic absorbers weighted by a total hydrogen absorption profile  \citep[c.f.][]{Paresce1980, Madau1995,Inoue2008,McCandliss2017},

\begin{equation}
<\tau(\lambda)>=\sum_{i=0}^{L}\sum_{j=0}^{M} \left(\frac{\partial^{2}n}{\partial{N_{HI}}\partial{z}}\right)_{i,j}(1-e^{-\tau(\lambda_{o})} )\Delta z_i  \Delta N_j,
\label{eqtrans}
\end{equation}
where the observed wavelength is related to the rest frame wavelength by $\lambda_{o} = \lambda_{r}(1+z_i)$.  The optical depth is $\tau(\lambda_{o}) = \tau_{HI}(\lambda_{o}) +\tau_{HeI}(\lambda_{o}) +\tau_{HeII}(\lambda_{o})$.  It includes all the resonance series lines from \ion{H}{1}, \ion{He}{1}, and \ion{He}{2} along with their respective opacities in the \lyc.  For \ion{H}{1} the LyC opacity is well approximated by a cross section that goes as $\propto (\lambda/\lambda_{edge})^3$ \citep{Spitzer1959} but is otherwise zero for $\lambda >$ 911.8 \AA, 
\begin{equation}
\sigma(\lambda)\approx 6.3 \times 10^{-18} (\lambda/911.8)^{3} cm^{2}.
\label{hicross}
\end{equation} 
The \ion{He}{2} cross section follows a roll-off with wavelength shortward of its ionization edge at 227.8 \AA\ that is the same as for \ion{H}{1} but 4 times smaller.  The \ion{He}{1} cross section shortward of 504.3 \AA\ is less extreme going as 7.4 $\times$ 10$^{-18} (\lambda/504.3)^{1.63}$ cm$^{2}$ \citep[c.f.][]{Shull2025, Verner1996, Samson1994}.

Corrections for the opacity of the intervening forest progressively increase with increasing redshift, although there exists the possibility of “lucky sightlines” as indicated by the height of the vertical crosses. The paucity of Lyman absorbers at low redshift elevates the importance of the Lyman UV (LUV) observer frame, between \lya\ and the Lyman edge, to quantify the ionizing radiation escape process. 

\begin{figure*}[t]
\centering
\includegraphics[width=.89\textwidth,clip,viewport=.8in 1.0in 8.2in 7.83in]{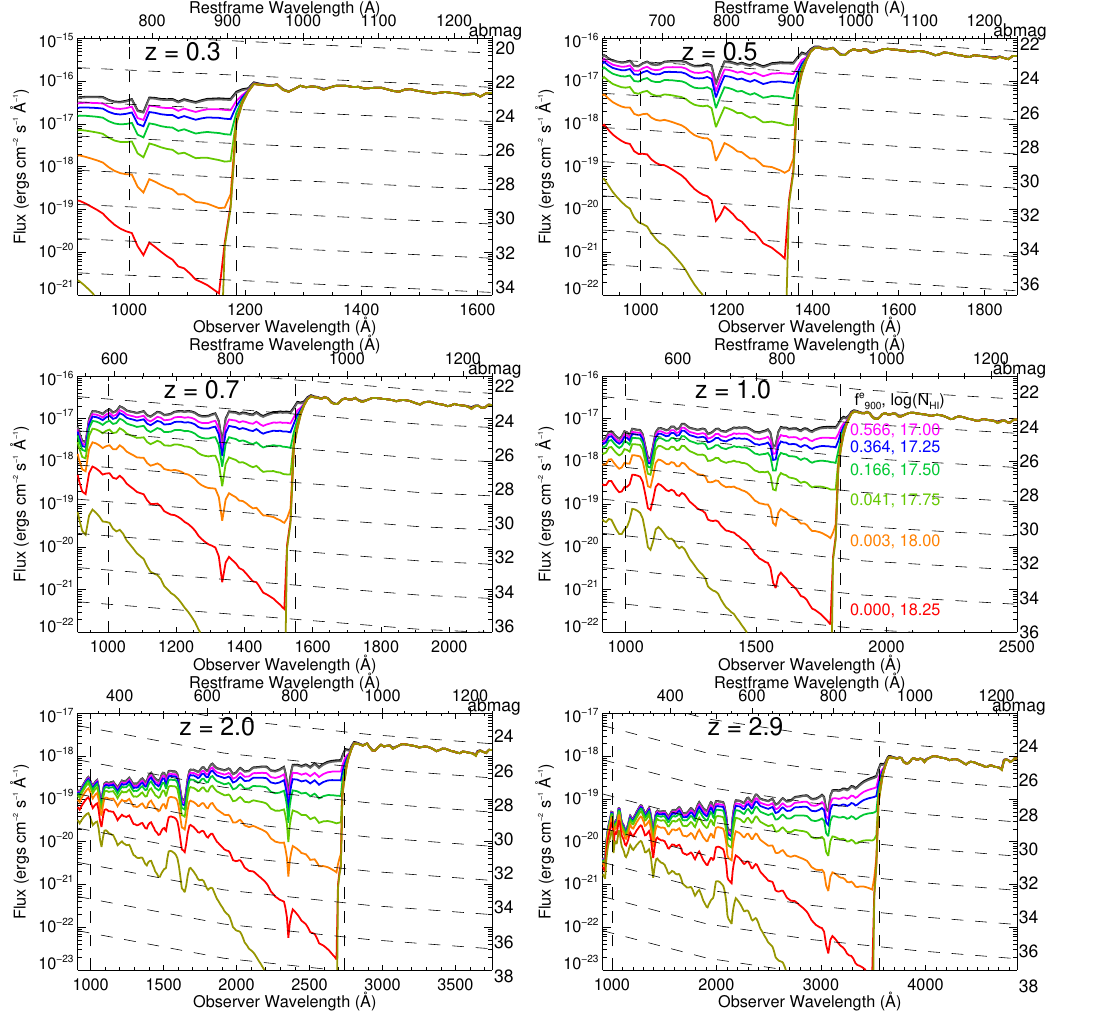}
\vspace{-.19in}
\caption{Redshifted attenuated SB99 models with logarithmic scaling, showing LyC drop-ins towards shorter wavelengths. Contours of constant abmag appear as dashed lines.  Escape fractions at 900 \AA\ ($f^e_{900}$ = 0.000, 0.000, 0.003, 0.041, 0.166, 0.364, 0.566, 0.945) correspond to column densities $\log{N_{HI}(cm^{-2})}$ = 18.50, 18.25, 18.00, 17.75, 17.50, 17.25, 17.00, 16.00); shown in olive, red, orange, light green, green, blue, violet, and grey respectively. The $z$ = 1 panel, middle-right, shows the  $f^e_{900}$ escape fractions and associated column densities. Lucky sightlines for $z \gtrsim$ 2 may offer \lyc\ detection.} 
\label{seds}
\vspace{-.19in}
\end{figure*}

The requirements for direct detection of LyC escape from low $z$ galaxies with a $L^{*}_{1500}$ luminosity are shown in Figure~\ref{seds} \citep{McCandliss2017} for escape fractions at the Lyman edge of \fen\ = (2e-8, 4e-5, 0.003, 0.041, 0.166, 0.364, 0.566, 0.945).  We see that the requirement for detection of a escape fraction at 912(1+z) \AA\ of $\sim$ 0.3\% is abmag = 30 or $\approx$ 4.5 $\times$ 10$^{-20}$ \flunit\ at a redshift of $z =$ 0.7 (orange line).

We note that to date no measurement of the actual shape of the SED below 912 \AA\ has ever been made.  The expectation is it will follow the shape of the intrinsic galactic SED as attenuated by a neutral hydrogen absorption cross-section, $\propto (\lambda/\lambda_{edge})^3$ -- neglecting order unity Gaunt factor variations with respect to wavelength  \citep{Gaunt1930, Karzas1961, Verner1996, Janicki1990}).  

However, recent calculations by \citet{Simmonds2024}, following \citet{Inoue2010, Inoue2011}, have shown that nebular emission in the form of free - bound transitions (free electrons recombining directly to $n =$ 1 level of hydrogen) can contribute to a flux excess just shortward of the ionization edge at  912 \AA, dubbed the Lyman bump.  The presence of a Lyman bump makes the use of the 900 \AA\ region to determine the fraction of ionizing photons that escape from star-forming galaxies problematic. 

Variations in dust opacity are also expected to produce attenuation deviations from $\lambda^3$ due to variations in the real and imaginary parts of the dielectic functions adopted for graphite and amorphous silicates \citep{Weingartner2001, Draine2003}.  The relative contribution of hard and soft ionizing sources to the aggregate stellar population (x-ray binaries, stripped stars, Wolf-Rayet stars, OB stars, alpha-enhanced, hot star binaries...) can also influence the intrinsic SED \citep{Hovis-Afflerbach2025, Byrne2025} as well as the ionized and neutral contributions of helium absorption to the attenuation.

\begin{figure*}
\centering
\includegraphics[width=.44\textwidth]{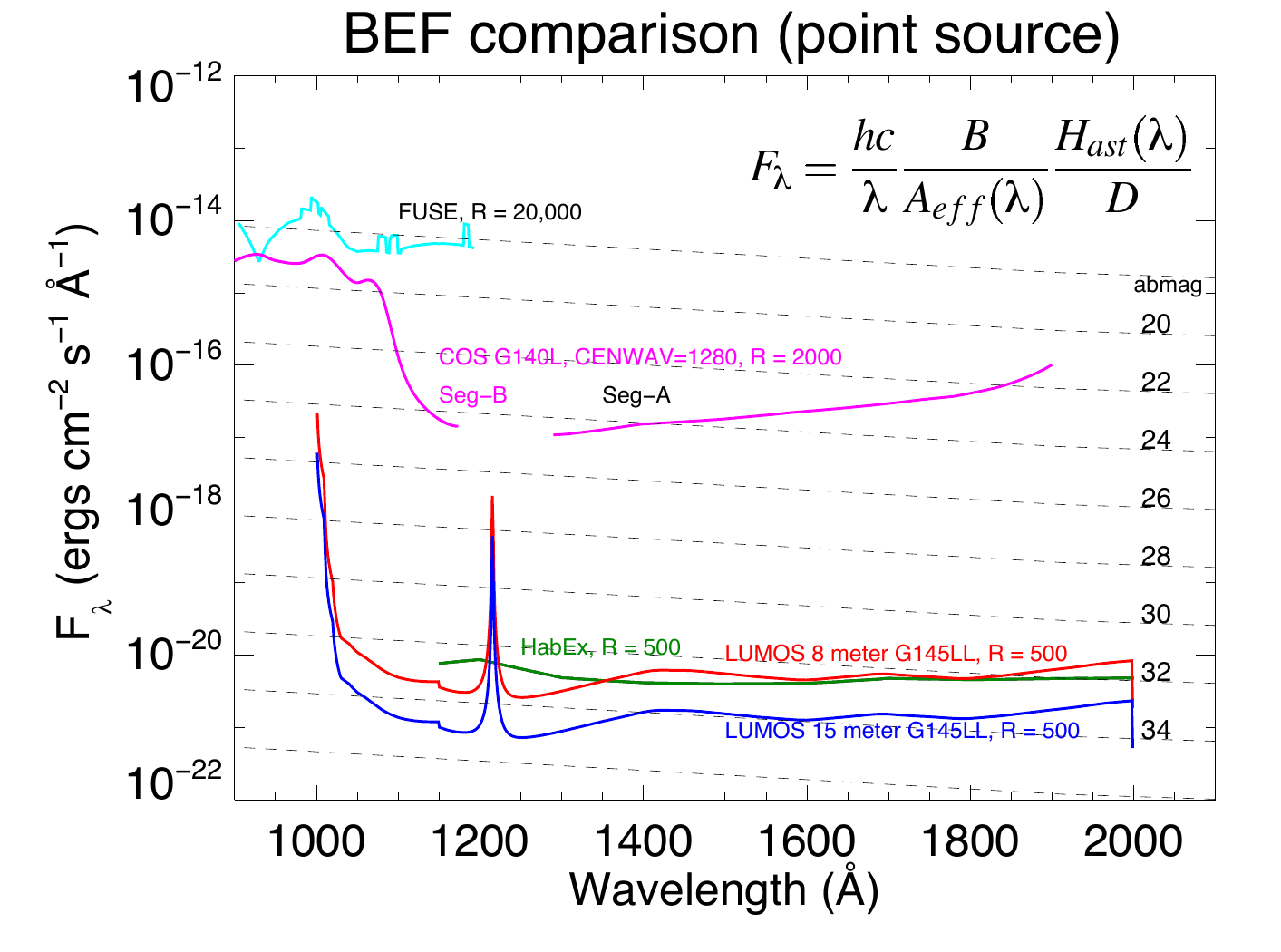}
\includegraphics[width=.54\textwidth,viewport=1in 0in 8.5in 4.5in,clip]{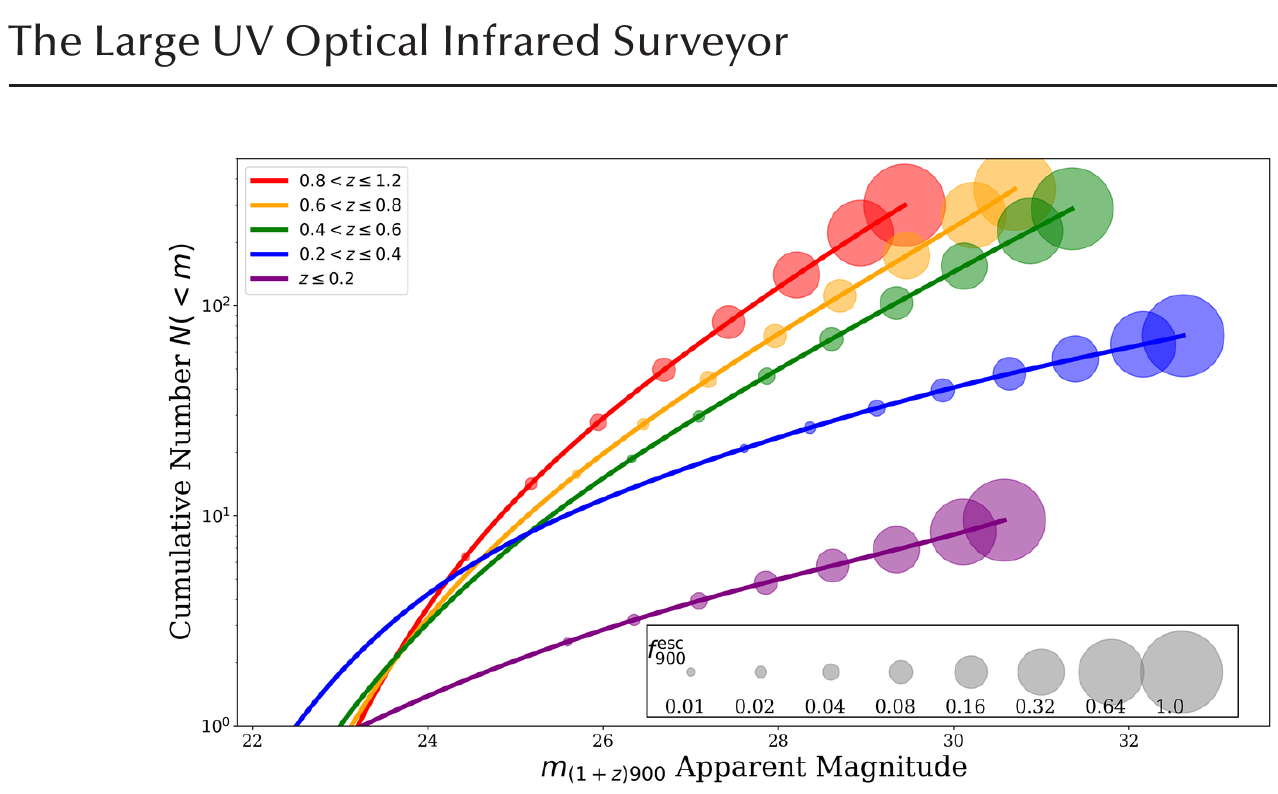}
\caption{ Left - Background Equivalent Flux (BEF)  from LUVOIR study \citep{LUVOIR2019}.  Right Cumulative detections in 2' $\times$ 2' FOV in 10 hours.} 
\label{bef}
\end{figure*}

These effects are important to account for because the escape of ionizing radiation measured just below the edge does not fully account for the  escape faction  integrated  over the whole \lyc\ emitting portion of a galaxy's SED. The integrated escape fraction (\fec) depends upon the detailed distribution of the neutral fractions of hydrogen  ($\chi_{HI} = \frac{N_{HI}}{N_{H}^{tot}}$), helium ($\chi_{HeI} = \frac{N_{HeI}}{N_{He}^{tot}}$), and dust variation in the galaxy's circumgalactic medium.  Elementary calculations show that it is possible for the ``edge'' escape fraction, \fen $<$ 1\%, while the  integrated \fec\ $\sim$ 10\%  \citep{McCandliss2017}.  Hence, a secondary objective of this study is to characterize the shape of the ionizing continuum below the hydrogen ionization edge to fully account for the sources of the MIB.


Understanding the process of reionization is key to understanding the transition of the universe from a predominantly neutral gas to a plasma state. Reionization is thought to have been facilitated by star-forming galaxies and active galactic nuclei, both of which produce escaping ionizing radiation at levels that are currently unconstrained \citep{Finkelstein2019, Madau2024}. How much radiation was produced, how much of it escaped the galaxies, and how escape was facilitated are questions that cannot be fully answered using measurements from the reionization era, as most of the ionizing UV radiation is absorbed by intervening neutral hydrogen along the line-of-sight. Galaxies in the local universe that have analogous properties to those sources in the reionization era will provide a unique and directly observable characterization of LyC escape. \\[-.25in]

\begin{figure*}
\centering
\includegraphics[width=\textwidth,viewport=1in 1.1in 6.5in 5.in,clip]{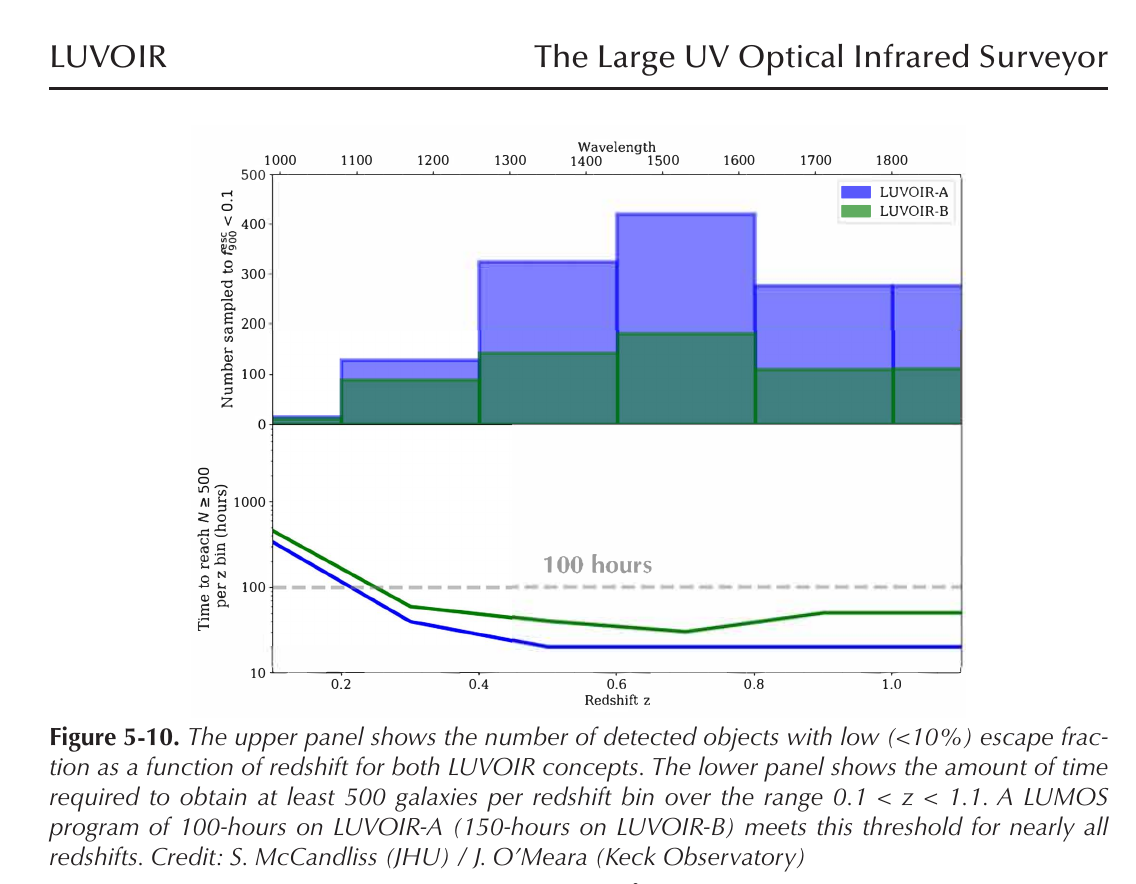}
\caption{ The upper panel shows the number of detected objects with low ($<$10\%) escape fraction as a function of redshift for both LUVOIR concepts. The lower panel shows the amount of time required to obtain at least 500 galaxies per redshift bin over the range 0.1 $< z <$ 1.1. A program of 150-hours with a LUVOIR-B like BEF meets this threshold for nearly all redshifts \citep{LUVOIR2019}.} 
\label{detects}
\end{figure*}

\begin{table*}
\caption{Physical Parameters}
\smallskip
\begin{center}
{\small
\begin{tabular}{|m{0.16\textwidth}|m{0.16\textwidth}|m{0.16\textwidth}|m{0.16\textwidth}|m{0.16\textwidth}|}  
\tableline
Physical Parameter & State of the Art &  Incremental Progress\pb(Enhancing) & Substantial Progress (Enabling) & Major Progress \pb(Breakthrough) \\
\tableline
Number of Galaxies (Ngal) 
& \raggedright LzLyC+, HST-15626 (PI  Jaskot)\pb\
 
 Ngal$ < $100 sampled at z $\sim$ 0.3 $\Delta z$ = 0.17 \pb\ 

Highly biased sample precludes robust determination of LyC luminosity function 
& Sprite cubesat (Fleming - PI) incremental progress to lower redshift  ($z \ge$ 0.15) \pb\

UVEX improved LF$_{1500(1+z)}$ functions out to $z$ = 1.2 and a large FoV (3\fd5)$^2$ and the potential to provide far-UV photometry of LyC beyond $z \sim$ 0.7   

&Ngal = 200 sampled per $\Delta z$ = 0.3 redshift bin out to $z$ = 1.0 or beyond, observed in a cosmic variance limited sample of $\sim$ (1$^{\circ}$)$^2$ 

&Ngal = 500  sampled per $\Delta z$ = 0.2 redshift bin out to $z$ = 1.2 or beyond, observed in a cosmic variance limited sample of $\sim$ (1$^{\circ}$)$^2$ \\
 \tableline

\fen 
 & 0.02 $\le$ fesc$\le$ 0.7 
 & fesc $\le$ 0.02
 & fesc $\le$ 0.01
 & fesc $\le$ 0.003\\
 
\tableline 
 Redshift Range
 & 0.26 $\le z \le$ 0.43 
 & 0.15 $\le z\lesssim$ 0.3
 & 0.2 $\le z\lesssim$ 1.0
 & 0.2 $\le z \lesssim$ 1.2 
 \\
 
 \tableline 
 LyC Lum Limit\pb(\lunit)
 & $\approx$ 2 $\times$ 10$^{39}$ 
 & $\approx$ 2 $\times$ 10$^{40}$ 
 & $\approx$ 1.3 $\times$ 10$^{38}$ 
 & $\approx$ 2 $\times$ 10$^{37}$ 
 \\
 
 \tableline
  abMags$_{900(1+z)}$ 
 & $\approx$ 25th 
 & 22.5 -- 22.3
 & 29.9 -- 28.6
 & 32.4 -- 31.1\\

\tableline
$F_{900(1+z)}$\pb(\flunit)  
 & 10$^{-17}$  
 & 10$^{-16}$
 & 10$^{-19}$
 & 10$^{-20}$ \\
\tableline
\end{tabular}
}\label{tab1}
\vspace{-.1in}
\end{center}
\end{table*}

\section{Physical Parameters}
Characterizing the production of ionizing radiation, along with the integrated escape fraction, from a diverse population of galaxies across cosmic time requires measurements of flux in the rest frame ionizing photon bandpass. By targeting galaxies with a range of physical properties we can identify the nature of ionizing sources and the physical conditions that allow the escape of ionizing radiation. These observations will provide an accurate determination of the sources and sinks of the ionization background radiation field at large. Most importantly, they will identify those ensembles of massive star populations that generate the ionizing radiation.  High spatial resolution could further isolate escape from individual clusters within a galaxy and offer insight to the geometric and kinematic properties of escape, however, such a science case has a narrow field-of-view requirements that are better addressed with integral field techniques rather than the wide field multi-object spectroscopic (MOS) capability required here.

The construction of a LyC luminosity function requires a flux limited sample of a large number of objects with known distances (redshifts) from which the luminosity can be calculated and volume distribution estimated.  Morphological information can be used to estimate intrinsic SEDs to allow determination of the fraction of LyC photons that escape from each individual galaxy (\fec). 

Characterization of the LyC luminosity function as a function of redshift requires statistically significant samples in bins of $\Delta z$ = 0.2, which are commensurate with those used to characterize the \GALEX\ $L_{1500(1+z)}$ luminosity function \citep{Arnouts2005}. Sample size per unit redshift interval and angular coverage are the important parameters to specify.  \GALEX\  acquired its sample in $\sim$ (1$^{\circ}$)$^{2}$.  We see from Figure~\ref{galex} that there are about 1000 $L^{*}$ galaxies per square degree at $z \sim$ 1.   A significant goal would be to sample $\sim$ 500 per redshift interval bin. Cosmic variance would be suppressed to $\sim$ 4\%; i.e. (500)$^{-\frac{1}{2}}$.  However, only a fraction of those galaxies sampled will actually exhibit LyC leakage.  Determining the faction that does is the raison d’\^{e}tre of this program.

The critical observational requirement will be the ability to detect ionizing radiation over the rest frame wavelength interval from 500 - 900 \AA.   From Figure~\ref{seds} we see that a background equivalent flux (BEF)  as deep as  (10$^{-20}$ \flunit) in the observer frame wavelength interval between 1000 - 1800 \AA\  (32.4 to 31.1 abmag respectively) would provide detection of an escape fraction 900 \AA\ to less than a fraction of a percent for $L^{*}$ galaxies, and offer insight into the shape of the ionizing photon spectral energy distribution in a wholly unexplored rest frame bandpass.  

The  Physical Parameters and for this program are summarized in Table~\ref{tab1} in various categories from current state of the art to breakthrough progress.

\begin{table*}
\caption{Observational Requirements}
\begin{center}
{\small
\begin{tabular}{|m{0.17\textwidth}|m{0.16\textwidth}|m{0.16\textwidth}|m{0.16\textwidth}|m{0.16\textwidth}|}  
\tableline
Observation \pb\ Requirement & State of the Art &  Incremental Progress\pb(Enhancing) & Substantial Progress (Enabling) & Major Progress \pb(Breakthrough) \\
\tableline
\raggedright MultiObject Spectroscopy
(HST/COS does not have MOS capability) 
& \raggedright Suborbital Rockets $A_{eff} \sim$ 50 cm$^2$ \hspace{.1in}
43 $\times$ 86 shutters
& \raggedright Smallsat Technology Accelerated Maturation Platform
128 $\times$ 256 shutters
&736 $\times$ 384 shutters
$A_{eff} \sim$ 80,000 cm$^2$
& 3 $\times$ (736 $\times$ 384) shut. 
$A_{eff} \sim$ 250,000 cm$^2$\\
 \tableline
Wavelength Range  
 & 1000 - 1800 \AA 
 & 1000 - 1800 \AA 
 & 1000 - 1800 \AA 
 & 1000 - 1800 \AA \\
 
 \tableline
Spectral Resolution 
 &  $R \approx $ 1000
 &  $R \approx $ 2000
 &  $R \approx $ 500
 &  $R \approx $ 1000\\
 
\tableline 
 Areal covered (FOV)
 & $(\frac{1}{2}^{\circ})^2$ 
 & $(\frac{1}{2}^{\circ})^2$  
 & (2\arcmin)$^2$ 
 & 3 $\times$ (2\arcmin)$^2$ 
 \\
 
\tableline 
 Field/Integration
 & Single field \pb\ 360 sec/field 
 & Single field \pb\ 86,400 sec/field
 & 15 fields \pb\ 36,000 sec/field
 & 15 fields \pb\ 36,000 sec/field
 \\

 \tableline 
 Target abMag
 & 15th at 1500 \AA
 & 17th at 1100 \AA
 & 30 at 900(1+z) \AA
 & 32 at 900(1+z) \AA
 \\
 
 \tableline
\end{tabular}
}\label{tab2}
\end{center}\vspace{-.3in}
\end{table*}

\section{Description of Observations}

In Figure~\ref{bef} left we provide BEF (left) and yield estimates (right) developed for the LUVOIR study, assuming a spectral resolution of $R$ = 500 and a field-of-view (FOV) of 2’ x 2’  in a 10 hour observation.  The size of the circles in the panel on the right indicated the level of escaping LyC that could be detected for an object with an apparent magnitude on the x-axis.

Figure~\ref{detects} shows the number that could be sampled to an escape fraction of less than 10\% in 150 hours for LUVOIR-B.  A BEF much with a limiting lower than 32 abmag would impact the number of LyC leakers that would be detected. Going forward the intent is to develop simulations that could answer the question of wiggle room for such a program. A lower limiting magnitude will require a reassessment of the program yield and the time required to acquire the statistically significant sample size.

The  Observational Requirements and for this program are summarized in Table~\ref{tab2} in various categories from current state of the art to breakthrough progress.

\section{Conclusion}
The objectives of this science case will transform our understanding of the physical processes that govern the \fec\ by obtaining the first spectrally resolved SEDs of LyC leakage.  They will provide the first critical insight into the nature of dust below the Lyman edge and the ionization fractions of hydrogen and helium in the halos of \lyc\ emitting galaxies, and how the meta-galactic radiation field is created and sustained over cosmic time.  It is an observing program that is most efficiently carried out with a MOS capability on a large HWO, the successful development of which will benefit a host of spectroscopic studies  beyond the \lyc\ and will be no less transformative.

{\bf Acknowledgements.} This work has been supported over the years by NASA grants to the Johns Hopkins University NNX08AM68G, NNX11AG54G, NNX17AC26G, 80NSSC22K0940, 80NSSC22K1698.  We also express thanks for useful ideas and lively conversations with Cody Carr, Anne Jaskot, Alberto Saldana-Lopez,  Claudia Scarlata, Mathew Hayes, John Chisholm, Daniel Schaerer, Timothy Heckman, Alaina Henry, Jack Ford, Brian Fleming, and Kevin France.
\bibliography{IonPhoLumFun4arXiv.bib}

\begin{thebibliography}{}
\parskip=0pt \itemsep=0pt \small \baselineskip=11pt
\expandafter\ifx\csname natexlab\endcsname\relax\def\natexlab#1{#1}\fi
\providecommand{\url}[1]{\href{#1}{#1}}
\providecommand{\dodoi}[1]{}
\providecommand{\doeprint}[1]{\href{http://ascl.net/#1}{#1}}
\providecommand{\doarXiv}[1]{\href{https://arxiv.org/abs/#1}{arXiv:#1}}

\bibitem[{{Alpher} {et~al.}(1948){Alpher}, {Bethe}, \& {Gamow}}]{Alpher1948}
{Alpher}, R.~A., {Bethe}, H., \& {Gamow}, G. 1948,
  \href{http://doi.org/10.1103/PhysRev.73.803}{\color{blue}Physical Review},
  \href{https://ui.adsabs.harvard.edu/abs/1948PhRv...73..803A}{\color{blue}73},
  803

\bibitem[{{Arnouts} {et~al.}(2005){Arnouts}, {Schiminovich}, {Ilbert},
  {Tresse}, {Milliard}, {Treyer}, {Bardelli}, {Budavari}, {Wyder}, {Zucca}, {Le
  F{\`e}vre}, {Martin}, {Vettolani}, {Adami}, {Arnaboldi}, {Barlow}, {Bianchi},
  {Bolzonella}, {Bottini}, {Byun}, {Cappi}, {Charlot}, {Contini}, {Donas},
  {Forster}, {Foucaud}, {Franzetti}, {Friedman}, {Garilli}, {Gavignaud},
  {Guzzo}, {Heckman}, {Hoopes}, {Iovino}, {Jelinsky}, {Le Brun}, {Lee},
  {Maccagni}, {Madore}, {Malina}, {Marano}, {Marinoni}, {McCracken}, {Mazure},
  {Meneux}, {Merighi}, {Morrissey}, {Neff}, {Paltani}, {Pell{\`o}}, {Picat},
  {Pollo}, {Pozzetti}, {Radovich}, {Rich}, {Scaramella}, {Scodeggio},
  {Seibert}, {Siegmund}, {Small}, {Szalay}, {Welsh}, {Xu}, {Zamorani}, \&
  {Zanichelli}}]{Arnouts2005}
{Arnouts}, S., {Schiminovich}, D., {Ilbert}, O., {et~al.} 2005,
  \href{http://doi.org/10.1086/426733}{\color{blue}\apjl},
  \href{http://adsabs.harvard.edu/abs/2005ApJ...619L..43A}{\color{blue}619},
  L43

\bibitem[{{Byrne} {et~al.}(2025){Byrne}, {Eldridge}, \& {Stanway}}]{Byrne2025}
{Byrne}, C.~M., {Eldridge}, J.~J., \& {Stanway}, E.~R. 2025,
  \href{http://doi.org/10.1093/mnras/staf178}{\color{blue}\mnras},
  \href{https://ui.adsabs.harvard.edu/abs/2025MNRAS.537.2433B}{\color{blue}537},
  2433

\bibitem[{{Draine}(2003)}]{Draine2003}
{Draine}, B.~T. 2003, \href{http://doi.org/10.1086/379118}{\color{blue}\apj},
  \href{https://ui.adsabs.harvard.edu/abs/2003ApJ...598.1017D}{\color{blue}598},
  1017

\bibitem[{{Fern{\'a}ndez-Soto} {et~al.}(2003){Fern{\'a}ndez-Soto}, {Lanzetta},
  \& {Chen}}]{Fernandez-Soto2003}
{Fern{\'a}ndez-Soto}, A., {Lanzetta}, K.~M., \& {Chen}, H.~W. 2003,
  \href{http://doi.org/10.1046/j.1365-8711.2003.06622.x}{\color{blue}\mnras},
  \href{https://ui.adsabs.harvard.edu/abs/2003MNRAS.342.1215F}{\color{blue}342},
  1215

\bibitem[{{Finkelstein} {et~al.}(2019){Finkelstein}, {D'Aloisio},
  {Paardekooper}, {Ryan}, {Behroozi}, {Finlator}, {Livermore}, {Upton
  Sanderbeck}, {Dalla Vecchia}, \& {Khochfar}}]{Finkelstein2019}
{Finkelstein}, S.~L., {D'Aloisio}, A., {Paardekooper}, J.-P., {et~al.} 2019,
  \href{http://doi.org/10.3847/1538-4357/ab1ea8}{\color{blue}\apj},
  \href{https://ui.adsabs.harvard.edu/abs/2019ApJ...879...36F}{\color{blue}879},
  36

\bibitem[{{Flury} {et~al.}(2022{\natexlab{a}}){Flury}, {Jaskot}, {Ferguson},
  {Worseck}, {Makan}, {Chisholm}, {Saldana-Lopez}, {Schaerer}, {McCandliss},
  {Wang}, {Ford}, {Heckman}, {Ji}, {Giavalisco}, {Amorin}, {Atek}, {Blaizot},
  {Borthakur}, {Carr}, {Castellano}, {Cristiani}, {De Barros}, {Dickinson},
  {Finkelstein}, {Fleming}, {Fontanot}, {Garel}, {Grazian}, {Hayes}, {Henry},
  {Mauerhofer}, {Micheva}, {Oey}, {Ostlin}, {Papovich}, {Pentericci},
  {Ravindranath}, {Rosdahl}, {Rutkowski}, {Santini}, {Scarlata}, {Teplitz},
  {Thuan}, {Trebitsch}, {Vanzella}, {Verhamme}, \& {Xu}}]{Flury2022a}
{Flury}, S.~R., {Jaskot}, A.~E., {Ferguson}, H.~C., {et~al.}
  2022{\natexlab{a}},
  \href{http://doi.org/10.3847/1538-4365/ac5331}{\color{blue}\apjs},
  \href{https://ui.adsabs.harvard.edu/abs/2022ApJS..260....1F}{\color{blue}260},
  1

\bibitem[{{Flury} {et~al.}(2022{\natexlab{b}}){Flury}, {Jaskot}, {Ferguson},
  {Worseck}, {Makan}, {Chisholm}, {Saldana-Lopez}, {Schaerer}, {McCandliss},
  {Xu}, {Wang}, {Oey}, {Ford}, {Heckman}, {Ji}, {Giavalisco}, {Amor{\'\i}n},
  {Atek}, {Blaizot}, {Borthakur}, {Carr}, {Castellano}, {De Barros},
  {Dickinson}, {Finkelstein}, {Fleming}, {Fontanot}, {Garel}, {Grazian},
  {Hayes}, {Henry}, {Mauerhofer}, {Micheva}, {Ostlin}, {Papovich},
  {Pentericci}, {Ravindranath}, {Rosdahl}, {Rutkowski}, {Santini}, {Scarlata},
  {Teplitz}, {Thuan}, {Trebitsch}, {Vanzella}, \& {Verhamme}}]{Flury2022b}
---. 2022{\natexlab{b}},
  \href{http://doi.org/10.3847/1538-4357/ac61e4}{\color{blue}\apj},
  \href{https://ui.adsabs.harvard.edu/abs/2022ApJ...930..126F}{\color{blue}930},
  126

\bibitem[{{Gamow}(1946)}]{Gamow1946}
{Gamow}, G. 1946,
  \href{http://doi.org/10.1103/PhysRev.70.572.2}{\color{blue}Physical Review},
  \href{https://ui.adsabs.harvard.edu/abs/1946PhRv...70..572G}{\color{blue}70},
  572

\bibitem[{{Gamow}(1948)}]{Gamow1948}
---. 1948, \href{http://doi.org/10.1038/162680a0}{\color{blue}\nat},
  \href{https://ui.adsabs.harvard.edu/abs/1948Natur.162..680G}{\color{blue}162},
  680

\bibitem[{{Gaunt}(1930)}]{Gaunt1930}
{Gaunt}, J.~A. 1930,
  \href{http://doi.org/10.1098/rspa.1930.0034}{\color{blue}Proceedings of the
  Royal Society of London Series A},
  \href{https://ui.adsabs.harvard.edu/abs/1930RSPSA.126..654G}{\color{blue}126},
  654

\bibitem[{{Gnedin} \& {Ostriker}(1997)}]{Gnedin1997}
{Gnedin}, N.~Y., \& {Ostriker}, J.~P. 1997,
  \href{http://doi.org/10.1086/304548}{\color{blue}\apj},
  \href{https://ui.adsabs.harvard.edu/abs/1997ApJ...486..581G}{\color{blue}486},
  581

\bibitem[{{Hovis-Afflerbach} {et~al.}(2025){Hovis-Afflerbach}, {G{\"o}tberg},
  {Schootemeijer}, {Klencki}, {Strom}, {Ludwig}, \&
  {Drout}}]{Hovis-Afflerbach2025}
{Hovis-Afflerbach}, B., {G{\"o}tberg}, Y., {Schootemeijer}, A., {et~al.} 2025,
  \href{http://doi.org/10.1051/0004-6361/202453185}{\color{blue}\aap},
  \href{https://ui.adsabs.harvard.edu/abs/2025A&A...697A.239H}{\color{blue}697},
  A239

\bibitem[{{Inoue}(2010)}]{Inoue2010}
{Inoue}, A.~K. 2010,
  \href{http://doi.org/10.1111/j.1365-2966.2009.15730.x}{\color{blue}\mnras},
  \href{https://ui.adsabs.harvard.edu/abs/2010MNRAS.401.1325I}{\color{blue}401},
  1325

\bibitem[{{Inoue} \& {Iwata}(2008)}]{Inoue2008}
{Inoue}, A.~K., \& {Iwata}, I. 2008,
  \href{http://doi.org/10.1111/j.1365-2966.2008.13350.x}{\color{blue}\mnras},
  \href{https://ui.adsabs.harvard.edu/abs/2008MNRAS.387.1681I}{\color{blue}387},
  1681

\bibitem[{{Inoue} {et~al.}(2014){Inoue}, {Shimizu}, {Iwata}, \&
  {Tanaka}}]{Inoue2014}
{Inoue}, A.~K., {Shimizu}, I., {Iwata}, I., {et~al.} 2014,
  \href{http://doi.org/10.1093/mnras/stu936}{\color{blue}\mnras},
  \href{http://adsabs.harvard.edu/abs/2014MNRAS.442.1805I}{\color{blue}442},
  1805

\bibitem[{{Inoue} {et~al.}(2011){Inoue}, {Kousai}, {Iwata}, {Matsuda},
  {Nakamura}, {Horie}, {Hayashino}, {Tapken}, {Akiyama}, {Noll}, {Yamada},
  {Burgarella}, \& {Nakamura}}]{Inoue2011}
{Inoue}, A.~K., {Kousai}, K., {Iwata}, I., {et~al.} 2011,
  \href{http://doi.org/10.1111/j.1365-2966.2010.17851.x}{\color{blue}\mnras},
  \href{https://ui.adsabs.harvard.edu/abs/2011MNRAS.411.2336I}{\color{blue}411},
  2336

\bibitem[{{Izotov} {et~al.}(2016{\natexlab{a}}){Izotov}, {Orlitov{\'a}},
  {Schaerer}, {Thuan}, {Verhamme}, {Guseva}, \& {Worseck}}]{Izotov2016a}
{Izotov}, Y.~I., {Orlitov{\'a}}, I., {Schaerer}, D., {et~al.}
  2016{\natexlab{a}},
  \href{http://doi.org/10.1038/nature16456}{\color{blue}\nat},
  \href{http://adsabs.harvard.edu/abs/2016Natur.529..178I}{\color{blue}529},
  178

\bibitem[{{Izotov} {et~al.}(2016{\natexlab{b}}){Izotov}, {Schaerer}, {Thuan},
  {Worseck}, {Guseva}, {Orlitov{\'a}}, \& {Verhamme}}]{Izotov2016b}
{Izotov}, Y.~I., {Schaerer}, D., {Thuan}, T.~X., {et~al.} 2016{\natexlab{b}},
  \href{http://doi.org/10.1093/mnras/stw1205}{\color{blue}\mnras},
  \href{http://adsabs.harvard.edu/abs/2016MNRAS.461.3683I}{\color{blue}461},
  3683

\bibitem[{{Izotov} {et~al.}(2018{\natexlab{a}}){Izotov}, {Schaerer}, {Worseck},
  {Guseva}, {Thuan}, {Verhamme}, {Orlitov{\'a}}, \& {Fricke}}]{Izotov2018a}
{Izotov}, Y.~I., {Schaerer}, D., {Worseck}, G., {et~al.} 2018{\natexlab{a}},
  \href{http://doi.org/10.1093/mnras/stx3115}{\color{blue}\mnras},
  \href{https://ui.adsabs.harvard.edu/abs/2018MNRAS.474.4514I}{\color{blue}474},
  4514

\bibitem[{{Izotov} {et~al.}(2018{\natexlab{b}}){Izotov}, {Worseck}, {Schaerer},
  {Guseva}, {Thuan}, {Fricke}, \& {Orlitov{\'a}}}]{Izotov2018b}
{Izotov}, Y.~I., {Worseck}, G., {Schaerer}, D., {et~al.} 2018{\natexlab{b}},
  \href{http://doi.org/10.1093/mnras/sty1378}{\color{blue}\mnras},
  \href{https://ui.adsabs.harvard.edu/abs/2018MNRAS.478.4851I}{\color{blue}478},
  4851

\bibitem[{{Janicki}(1990)}]{Janicki1990}
{Janicki}, C. 1990,
  \href{http://doi.org/10.1016/0010-4655(90)90027-X}{\color{blue}Computer
  Physics Communications},
  \href{https://ui.adsabs.harvard.edu/abs/1990CoPhC..60..281J}{\color{blue}60},
  281

\bibitem[{{Karzas} \& {Latter}(1961)}]{Karzas1961}
{Karzas}, W.~J., \& {Latter}, R. 1961,
  \href{http://doi.org/10.1086/190063}{\color{blue}\apjs},
  \href{https://ui.adsabs.harvard.edu/abs/1961ApJS....6..167K}{\color{blue}6},
  167

\bibitem[{{Kashlinsky} \& {Rees}(1983)}]{Kashlinsky1983}
{Kashlinsky}, A., \& {Rees}, M.~J. 1983,
  \href{http://doi.org/10.1093/mnras/205.4.955}{\color{blue}\mnras},
  \href{https://ui.adsabs.harvard.edu/abs/1983MNRAS.205..955K}{\color{blue}205},
  955

\bibitem[{{Madau}(1995)}]{Madau1995}
{Madau}, P. 1995, \href{http://doi.org/10.1086/175332}{\color{blue}\apj},
  \href{http://adsabs.harvard.edu/abs/1995ApJ...441...18M}{\color{blue}441}, 18

\bibitem[{{Madau} {et~al.}(2024){Madau}, {Giallongo}, {Grazian}, \&
  {Haardt}}]{Madau2024}
{Madau}, P., {Giallongo}, E., {Grazian}, A., {et~al.} 2024,
  \href{http://doi.org/10.3847/1538-4357/ad5ce8}{\color{blue}\apj},
  \href{https://ui.adsabs.harvard.edu/abs/2024ApJ...971...75M}{\color{blue}971},
  75

\bibitem[{{McCandliss} \& {O'Meara}(2017)}]{McCandliss2017}
{McCandliss}, S.~R., \& {O'Meara}, J.~M. 2017,
  \href{http://doi.org/10.3847/1538-4357/aa7fbb}{\color{blue}\apj},
  \href{https://ui.adsabs.harvard.edu/abs/2017ApJ...845..111M}{\color{blue}845},
  111

\bibitem[{{Paresce} {et~al.}(1980){Paresce}, {McKee}, \&
  {Bowyer}}]{Paresce1980}
{Paresce}, F., {McKee}, C.~F., \& {Bowyer}, S. 1980,
  \href{http://doi.org/10.1086/158244}{\color{blue}\apj},
  \href{http://adsabs.harvard.edu/abs/1980ApJ...240..387P}{\color{blue}240},
  387

\bibitem[{{Redwine} {et~al.}(2016){Redwine}, {McCandliss}, {Zheng}, {Fleming},
  {France}, {Osterman}, {Howk}, {Anderson}, \& {G{\"a}ensicke}}]{Redwine2016}
{Redwine}, K., {McCandliss}, S.~R., {Zheng}, W., {et~al.} 2016,
  \href{http://doi.org/10.1088/1538-3873/128/968/105006}{\color{blue}\pasp},
  \href{https://ui.adsabs.harvard.edu/abs/2016PASP..128j5006R}{\color{blue}128},
  105006

\bibitem[{{Samson} {et~al.}(1994){Samson}, {He}, {Yin}, \&
  {Haddad}}]{Samson1994}
{Samson}, J.~A.~R., {He}, Z.~X., {Yin}, L., {et~al.} 1994,
  \href{http://doi.org/10.1088/0953-4075/27/5/008}{\color{blue}Journal of
  Physics B Atomic Molecular Physics},
  \href{https://ui.adsabs.harvard.edu/abs/1994JPhB...27..887S}{\color{blue}27},
  887

\bibitem[{{Schechter}(1976)}]{Schechter1976}
{Schechter}, P. 1976, \href{http://doi.org/10.1086/154079}{\color{blue}\apj},
  \href{https://ui.adsabs.harvard.edu/abs/1976ApJ...203..297S}{\color{blue}203},
  297

\bibitem[{{Shull} {et~al.}(2025){Shull}, {Curran}, \& {Topping}}]{Shull2025}
{Shull}, J.~M., {Curran}, R.~M., \& {Topping}, M.~W. 2025,
  \href{http://doi.org/10.3847/1538-4357/ad9b8b}{\color{blue}\apj},
  \href{https://ui.adsabs.harvard.edu/abs/2025ApJ...979...21S}{\color{blue}979},
  21

\bibitem[{{Silk}(1983)}]{Silk1983}
{Silk}, J. 1983,
  \href{http://doi.org/10.1093/mnras/205.3.705}{\color{blue}\mnras},
  \href{https://ui.adsabs.harvard.edu/abs/1983MNRAS.205..705S}{\color{blue}205},
  705

\bibitem[{{Simmonds} {et~al.}(2024){Simmonds}, {Verhamme}, {Inoue}, {Katz},
  {Garel}, \& {De Barros}}]{Simmonds2024}
{Simmonds}, C., {Verhamme}, A., {Inoue}, A.~K., {et~al.} 2024,
  \href{http://doi.org/10.1093/mnras/stae1003}{\color{blue}\mnras},
  \href{https://ui.adsabs.harvard.edu/abs/2024MNRAS.530.2133S}{\color{blue}530},
  2133

\bibitem[{{Spitzer} \& {Zabriskie}(1959)}]{Spitzer1959}
{Spitzer}, Jr., L., \& {Zabriskie}, F.~R. 1959,
  \href{http://doi.org/10.1086/127416}{\color{blue}\pasp},
  \href{https://ui.adsabs.harvard.edu/abs/1959PASP...71..412S}{\color{blue}71},
  412

\bibitem[{{The LUVOIR Team}(2019)}]{LUVOIR2019}
{The LUVOIR Team}. 2019

\bibitem[{{Verner} {et~al.}(1996){Verner}, {Ferland}, {Korista}, \&
  {Yakovlev}}]{Verner1996}
{Verner}, D.~A., {Ferland}, G.~J., {Korista}, K.~T., {et~al.} 1996,
  \href{http://doi.org/10.1086/177435}{\color{blue}\apj},
  \href{http://adsabs.harvard.edu/abs/1996ApJ...465..487V}{\color{blue}465},
  487

\bibitem[{{Wang} {et~al.}(2019){Wang}, {Heckman}, {Leitherer}, {Alexandroff},
  {Borthakur}, \& {Overzier}}]{Wang2019}
{Wang}, B., {Heckman}, T.~M., {Leitherer}, C., {et~al.} 2019,
  \href{http://doi.org/10.3847/1538-4357/ab418f}{\color{blue}\apj},
  \href{https://ui.adsabs.harvard.edu/abs/2019ApJ...885...57W}{\color{blue}885},
  57

\bibitem[{{Weingartner} \& {Draine}(2001)}]{Weingartner2001}
{Weingartner}, J.~C., \& {Draine}, B.~T. 2001,
  \href{http://doi.org/10.1086/318651}{\color{blue}\apj},
  \href{https://ui.adsabs.harvard.edu/abs/2001ApJ...548..296W}{\color{blue}548},
  296

\end{thebibliography}

\end{document}